\documentclass[prb,twocolumn,superscriptaddress,amsmath,amsfonts]{revtex4-2}

\usepackage{amsmath}
\usepackage{amssymb}
\usepackage{mathrsfs}
\usepackage{physics}
\usepackage{hyperref}
\usepackage[capitalize]{cleveref}
\usepackage{graphicx}
\usepackage{dcolumn}

\newcolumntype{d}[1]{D{.}{.}{#1}}
\newcommand\mc[1]{\multicolumn{1}{c}{#1}}

\begin{document}

\title{Separate measurement- and feedback-driven entanglement transitions in the stochastic control of chaos}
\author{Conner LeMaire}
\affiliation{Department of Physics and Astronomy, Louisiana State University, Baton Rouge, LA 70803, USA}
\author{Andrew A. Allocca}
\affiliation{Department of Physics and Astronomy, Louisiana State University, Baton Rouge, LA 70803, USA}
\affiliation{Center for Computation and Technology, Louisiana State University, Baton Rouge, LA 70803, USA}
\author{J. H. Pixley}
\affiliation{Department of Physics and Astronomy, Center for Materials Theory, Rutgers University, Piscataway, NJ 08854 USA}
\affiliation{
Center for Computational Quantum Physics, Flatiron Institute, 162 5th Avenue, New York, NY 10010
}%
\author{Thomas Iadecola}
\affiliation{Department of Physics and Astronomy, Iowa State University, Ames, IA 50011, USA}
\affiliation{Ames National Laboratory, Ames, IA 50011, USA}
\author{Justin H. Wilson}
\affiliation{Department of Physics and Astronomy, Louisiana State University, Baton Rouge, LA 70803, USA}
\affiliation{Center for Computation and Technology, Louisiana State University, Baton Rouge, LA 70803, USA}
\date{\today}

\begin{abstract}
We study measurement-induced entanglement and control phase transitions in a quantum analog of the Bernoulli map subjected to a classically-inspired control protocol.
When entangling gates are restricted to the Clifford group, separate entanglement ($p_\mathrm{ent}$) and control ($p_\mathrm{ctrl}$) transitions emerge, revealing two distinct universality classes.
The control transition has critical exponents $\nu$ and $z$ consistent
with the classical map (a random walk) while the entanglement transition is revealed to have similar exponents as the measurement-induced phase transition in Clifford hybrid dynamics.
This is distinct from the case of generic entangling gates in the same model, where $p_\mathrm{ent} = p_\mathrm{ctrl}$ and universality is controlled by the random walk.
\end{abstract}

\maketitle

\section{Introduction}

The collective dynamics of a chaotic many-body quantum system are difficult to predict and control.
Nonetheless, as many noisy intermediate-scale quantum devices come online, it is important to find robust ways of steering the dynamics to desired states.
Remarkably, this complex quantum control problem has an analog in the study of classical chaos.
Within that literature, a long-standing problem is how to steer chaotic dynamics onto unstable periodic orbits~\cite{Ott1990}.
In Refs.~\cite{Antoniou1996,Antoniou1997,Antoniou1998}, classical chaotic dynamics were controlled onto unstable periodic orbits by randomly measuring the system and performing a feedback operation to help steer the dynamics.
In these works, steering is achieved via random interventions without continuously monitoring the system.
This allows the problem to be analyzed within the framework of statistical physics where the regimes of control emerge from phase transitions in the dynamics of the system.

A similar type of nonequilibrium phase transition in the dynamics of quantum many-body systems has been extensively studied: the measurement-induced phase transition (MIPT) \cite{PotterVasseur2022a,FisherVijay2023}.
In the standard formulation, this transition is characterized by a change in the character of the steady-state many-body wave function~\cite{Skinner2019,Vasseur2019,Bao2020}.
Up to some critical rate of measurements $p_\mathrm{ent}$, the system is volume-law entangled (i.e., the reduced density matrix for a subsystem $A$ has an entanglement entropy proportional to its volume).
However, for rates $p > p_\mathrm{ent}$, the system's wave function becomes area-law entangled (i.e., the reduced density matrix over $A$ has entanglement entropy proportional to the boundary of $A$). 
In previous work~\cite{Iadecola2023}, some of us showed how these types of transitions could be unified with control transitions in a quantum version of the classically chaotic Bernoulli map~\cite{Renyi1957}. 
Other works have since appeared finding similar transitions (sometimes referred to as absorbing state transitions) in a wide variety of dynamics~\cite{Buchhold2022,MilekhinPopov2023,Friedman2022a,Ravindranath2022, ODea2022, SierantTurkeshi2023,SierantTurkeshi2023a}.
Interestingly, these control transitions are described by distinct universality classes: while the quantum Bernoulli map's criticality is described by a random walk \cite{Iadecola2023}, absorbing state transitions are consistent with the directed percolation universality class \cite{Hinrichsen2000,Odor2004}.

Despite the distinct nature of these control transitions, we can still identify some common features of all these models.
The first is that generically the rate of control operations involving measurement and feedback ($p_\mathrm{ctrl}$) must be greater than or equal to the rate of pure measurements that would drive an entanglement transition ($p_\mathrm{ent}$).
The reasoning for this comes from viewing the entanglement transition as a purification transition~\cite{Gullans2020a,Gullans2020b}; for $p< p_\mathrm{ent}$, quantum information is hidden from measurements for a time scaling exponentially with system size.
Since measurements cannot access this information, feedback operations cannot direct the whole system to a desired state.
However, once $p>p_\mathrm{ent}$ measurements are extracting that information, allowing one to potentially control the whole system.
Additionally, it appears that generic chaotic dynamics can saturate this bound so that $p_\mathrm{ctrl}=p_\mathrm{ent}$.
When this occurs, it appears as though the control universality ``wins''; it is still an open question as to why the control transition universality is dominant in trajectory dynamics.

\begin{figure}
    \centering
    \includegraphics[width=\columnwidth]{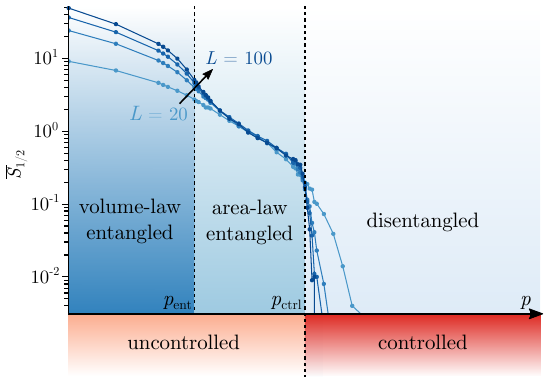}
    \caption{A phase diagram of the stabilizer version of the quantum Bernoulli map with control.
    The use of Clifford gates separates the entanglement ($p_\mathrm{ent}$) and control ($p_\mathrm{ctrl}$) transitions, giving three distinct phases. 
    At long times the average half-cut entanglement entropy, shown on a log scale, achieves a value that scales with the system size for $p<p_\mathrm{ent}$ (volume-law), is $O(1)$ for $p_\mathrm{ent}<p<p_\mathrm{ctrl}$ (area-law), and drops to zero for $p>p_\mathrm{ctrl}$ (disentangled).
    The entanglement transition exhibits universal features of the stabilizer MIPT~\cite{li_measurement-driven_2019,zabalo_critical_2020} while the control transition's criticality is derived from a random walk~\cite{Iadecola2023}.}
    \label{fig:phases}
\end{figure}

Building on our previous work in Ref.~\cite{Iadecola2023}, here we explore a stabilizer limit of the quantum Bernoulli map subject to stochastic control \cite{antoniou_probabilistic_1996}. 
This naturally modifies the dynamics away from the generic case studied in~Ref.~\cite{Iadecola2023}, where any state in the Hilbert space is accessible with sufficient repeated applications of the chaotic map.
The allowable states are now a discrete set of stabilizer states accessed by Clifford gates \cite{nielsen_quantum_2011} and, importantly, the control protocol is also implemented with Clifford operations, leaving the final state a stabilizer state even after the combined unitary dynamics and control.
This has the notable benefit that classical simulations can be performed to access much larger system sizes due to the Gottesman-Knill theorem \cite{aaronson_improved_2004}. 
Restricting the dynamics from the full Hilbert space to only a discrete subspace separates the two phase transitions---the control transition at $p=p_\mathrm{ctrl}$ described by a random walk, and the entanglement transition at $p=p_\mathrm{ent}<p_\mathrm{ctrl}$ described by a logarithmic conformal field theory. 
Thus, we find three separate phases and two distinct transitions: An uncontrolled volume-law-entangled phase, an uncontrolled area-law-entangled phase, and a controlled disentangled phase, as shown schematically in \cref{fig:phases}.
Notably, up to numerical accuracy the entanglement transition has certain critical exponents matching those of the stabilizer MIPT~\cite{li_quantum_2018, Gullans2020a,zabalo_critical_2020}, while the control transition inherits universal features from the classical transition~\cite{Iadecola2023,Antoniou1996}.
Recovering universal features of the stabilizer MIPT in addition to a separate control transition is one of the main results of this work.

The rest of the paper is organized as follows.
In Sec.~\ref{sec:model}, we review the Bernoulli map model and discuss modifications of the scheme presented in \cite{Iadecola2023} to allow stabilizer simulations without sacrificing the universal features of control.
In Sec.~\ref{sec:ctrl_transition} we confirm that the control transition persists and is captured by a classical symmetry-breaking order parameter.
On the other hand, we report the observation of a distinct entanglement transition in Sec.~\ref{sec:ent_transition} occurring before the control transition.
Finally, we end with a discussion in Sec.~\ref{sec:discussion}.

\section{Model and Observables}\label{sec:model}

As in Ref.~\cite{Iadecola2023}, we start from the classically-chaotic Bernoulli map $B$, which acts on points in the interval $x\in[0,1)$ as~\cite{Renyi1957}
\begin{equation}
    x \mapsto 2x \mod 1.
\end{equation}
The point $x$ can be represented as a binary fraction of infinite length, 
\begin{equation}
    x = 0.b_1 b_2 b_3 \cdots
\end{equation} 
with $b_i \in \{0,1\}$, such that $B$ simply shifts all bits to the left and deletes the first digit 
\begin{equation}
    0.b_1 b_2 b_3 \cdots \mapsto 0.b_2 b_3 b_4 \cdots.
\end{equation}
This map has periodic orbits if it is initialized with any rational number $x_0 = p/q$ for integers $p$ and $q$ with $q$ odd~\footnote{This ensures that the binary expansion does not terminate and have leading 0s that do not repeat.}.
Stochastic control will work on any of these orbits \cite{Antoniou1996}.
The simplest periodic orbit is the unstable stationary point $x_0=0$; universality dictates that control onto this fixed point has the same critical properties as control onto any other orbit.
Specifying $x_0=0$ as the target state for control gives rise to a simple control map $C$:
\begin{equation}
    x \mapsto x/2.
\end{equation}
The action of $C$ on a binary fraction shifts all bits to the right and adds a leading zero
\begin{equation}
    0.b_1 b_2 b_3 \cdots \mapsto 0.0 b_1 b_2 \cdots.
\end{equation}
The dynamics then proceed stochastically: with probability $p$ we apply $C$ and with probability $1-p$ we apply $B$.
In this particular setup, the control transition occurs at $p_\mathrm{ctrl}=1/2$~\cite{antoniou_absolute_1998}; we will find this is unaltered by the modified stabilizer dynamics below.
This implementation also makes manifest the object which governs the control transition: the position of the first $1$ in the bitstring, which we call the \textit{first domain wall}. $C$ moves the domain wall deeper into the binary expansion and $B$ moves it leftward towards the more significant digits.

\begin{figure}
    \centering
    \includegraphics[width=\columnwidth]{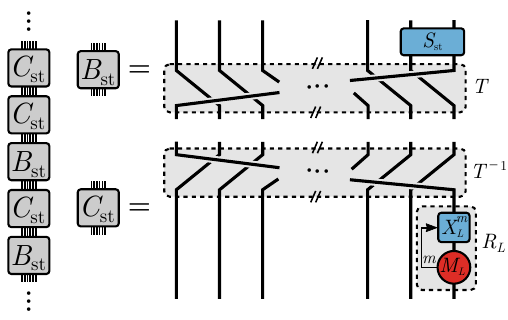}
    \caption{A schematic diagram of the circuit we consider and of the individual Bernoulli $B_\mathrm{st}$ and control $C_\mathrm{st}$ maps.
    In the Bernoulli map we first apply $T$ to bring the first qubit to position $L$, then apply a random 2-qubit Clifford gate $S_\mathrm{st}$ to scramble the least-significant digits.
    In the control map we first apply a reset operation $R_L$ on the final qubit, comprised of a measurement (the red circle labeled $M_L$) and a gate dependent on the measurement result $m$ (blue square labeled $X^m_L$).
    If $m=0$ we do nothing, and if $m=1$ we apply $X_L$. 
    We then apply $T^{-1}$ to move this final qubit, controlled onto the state $\ket{0}$, to the first position.}
    \label{fig:schematic}
\end{figure}

Following Ref.~\cite{Iadecola2023}, the extension of this map to the quantum case replaces these classical bits with qubits, so that a computational basis state of the system can be written as $\ket{x} = \ket{b_1 b_2 b_3\dots}$ and a general state as a superposition over these, $\ket{\psi} = \sum_x \psi_x \ket{x}$.
To allow for simulation, we truncate the system to $L$ qubits, giving a Hilbert space of size $2^L$.
The Bernoulli map is then given by the unitary operation
\begin{equation}\label{eq:Bst}
    B_\mathrm{st} = S_\mathrm{st} T,
\end{equation}
where $T$ is the translation operator $$T\ket{b_1 b_2 \dots b_L}=\ket{b_2b_3 \dots b_Lb_1 },$$ and $S_\mathrm{st}$ is a random 2-qubit Clifford gate acting on the last two qubits which acts as a ``scrambler'' (pictured in \cref{fig:schematic}).
In Ref.~\cite{Iadecola2023}, $S_\mathrm{st}$ was implemented with either a classical cellular automaton (i.e., a local permutation matrix) or a generic quantum (Haar) gate.
With stabilizer states, we still maintain the cellular automaton as a limit, but we no longer explore the full Hilbert space; instead, we explore the full subspace of stabilizer states if we choose $S_\mathrm{st}$ randomly.
In this sense, by repeated application of the $B_\mathrm{st}$ we explore all stabilizer states and are therefore ergodic on that subspace; however, we are restricted in the entanglement structure which is available in the generic case.

To understand the connection between stabilizers and classical cellular automata, note that the state $\ket{b_1 b_2}$ is \emph{stabilized} by two Pauli operators, $(-1)^{b_1} Z_1$ and $(-1)^{b_2} Z_2$, which uniquely specify the state \cite{nielsen_quantum_2011}.
A random stabilizer $S_\mathrm{st}$ can transform one of these Pauli operators into a different one (e.g., $S_\mathrm{st} Z_1 S_\mathrm{st}^{-1} = X_1 Y_2$) which in turn tells us how the state transforms as well.
A random 2-bit cellular automaton is a special case of this where $S_\mathrm{st} Z_i S_\mathrm{st}^{-1}=\pm Z_i$.
By keeping track of the stabilizers, we can simulate larger system sizes than the fully quantum case \cite{aaronson_improved_2004}.

In the computational basis representation, $x_0 = 0$ is the polarized state $\ket{00\cdots 0}$; this is in contrast to Ref.~\cite{Iadecola2023}, which targets the two state orbit containing $x_0 = 1/3$ and $x_1 = 2/3$, corresponding to N\'eel states in the binary representation. 
Control onto the period-2 orbit requires a control protocol that includes a full adder, which cannot generically be implemented with Clifford gates~\cite{nielsen_quantum_2011}.

To implement the control map $C_\mathrm{st}$ it is necessary to break unitarity and implement feedback.
The operator itself is
\begin{equation}
\label{eq:Cst}
    C_\mathrm{st} = T^{-1} R_L,
\end{equation}
where $R_L$ is the reset operation on the last qubit which first measures the last bit in the bitstring, $b_L$, and resets it to 0 if $\ket{b_L} = \ket{1}$ is measured
\begin{equation}
    R_{L}\ket{\psi} = \begin{cases}
        \frac{P_L^0 \ket{\psi}}{\lvert \lvert P_L^0 \ket{\psi}\rvert\rvert} & \text{with probability } \lvert \lvert P_L^0 \ket{\psi}\rvert\rvert^2, \\
        \frac{X_L P_L^1 \ket{\psi}}{\lvert \lvert P_L^1 \ket{\psi}\rvert\rvert} & \text{with probability } \lvert \lvert P_L^1 \ket{\psi}\rvert\rvert^2,
    \end{cases}
\end{equation}
where $P_L^b = (1 + (-1)^b Z_L)/2$ is the projector onto the measured outcome of the last qubit.
The operator $R_L$ is the crucial step of implementing measurement and feedback as seen in \cref{fig:schematic}.

Returning to the first domain wall picture, it is convenient to define the first domain wall's position in the Clifford setup via the expression 
\begin{equation}
\ket{\Psi} = \lvert\underbrace{00 \ldots 0}_{\ell}\rangle \otimes \ket{\psi}, \label{eq:fdw}
\end{equation}
where $\ell$ 0's precede a domain wall defined by $\ket{\psi}$ having a finite probability to have a 1 on the $\ell+1$ bit~\footnote{Note that this quantum generalization differs from the first domain wall definition discussed in Ref.~\cite{Iadecola2023}}.
In general, $\ell \mapsto \ell - 1$ under $B_\mathrm{st}$ and $\ell \mapsto \ell + 1$ under $C_\mathrm{st}$. 
This object will be useful for discussion in Sec.~\ref{sec:discussion}.

To simulate the Clifford circuit constructed with these maps, we use the Python package stim (version 1.9.0) \cite{gidney2021stim}, which allows for Clifford gate operations, measurements, and varying system sizes. 
We first initialize an $L$ qubit system in a random stabilizer state.
The dynamics are governed by the control probability $p$:
With probability $p$ we apply a control map, and with probability $1-p$ we apply the Bernoulli map.
This is repeated for $2L^2$ time steps, long enough for the system to reach a steady state, at which point the system's entanglement entropy can be evaluated using the stabilizer tableau. 

We will consider two different types of entanglement entropy, both of which can be analyzed as von Neumann entropies
\begin{equation}
    S_{A} = -\Tr[\rho_A\log_2\rho_A],
\end{equation}
where $A$ is some subset of the total system, and $\rho_A = \Tr_{\bar{A}}(\rho)$ is the reduced density matrix of subsystem $A$ after tracing over its complement $\bar{A}$.
First is the half-cut entropy $S_{1/2}$ measuring the degree of entanglement between two disjoint halves of the system, so that $A$ is comprised of qubits $1$ to $L/2$.
We also examine entanglement through the lens of purification: by maximally entangling an additional ancilla qubit with the initial state of the $L$ qubit system we can extract information about the system's purity by calculating the entanglement entropy of this ancilla with the system, $S_a$. 
In this case, subsystem $A$ is the ancilla itself. 
For both cases, the stabilizer formalism allows the calculation of the entropy \cite{HammaZanardi2005,HammaZanardi2005a,NahumHaah2017a}.
In our discussions of these entropies and other quantities below, we consider their values averaged over many realizations of our probabilistic circuit, indicated with a bar $\overline{(\cdots)}$ and all error bars, unless specified otherwise, are standard errors. 

\begin{figure*}[!ht]
    \centering
    \includegraphics[width=0.32\textwidth]{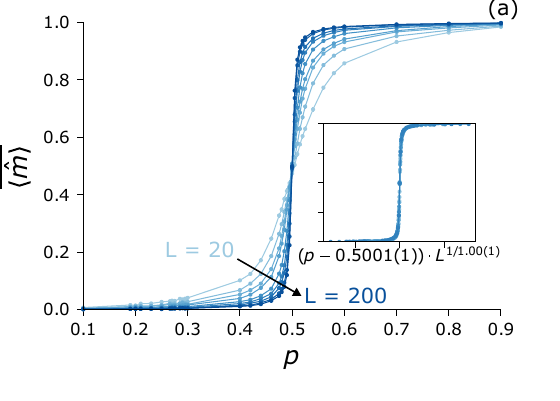}
    \includegraphics[width=0.32\textwidth]{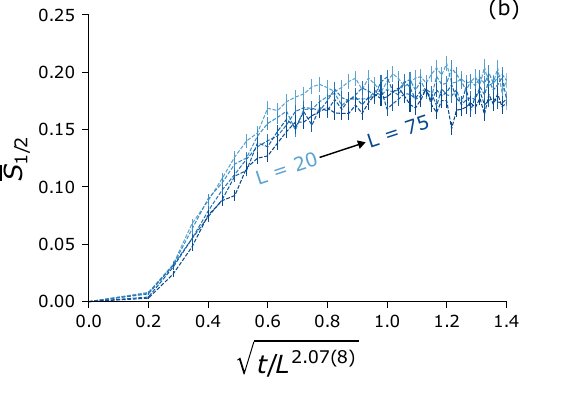}
    \includegraphics[width=0.32\textwidth]{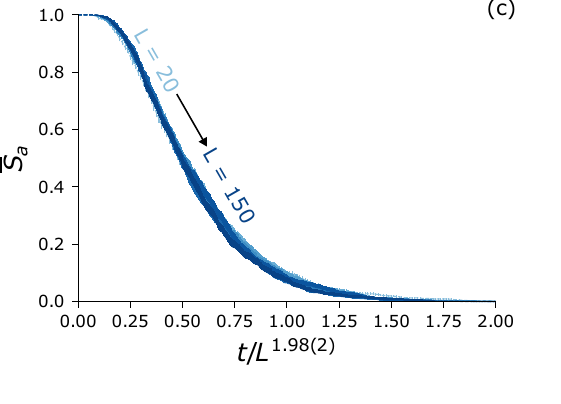}
    \caption{{\bf The Control Transition}:  (a) The magnetization density at $t=2L^2$ as a function of $p$. The inset shows data collapse indicating $p_\mathrm{ctrl} = 0.5001(1)$ and $\nu = 1.00(1)$. 
    (b) The dynamics of the half-cut entropy $\overline{S}_{1/2}$ and (c) the dynamics of the ancilla entropy $\overline{S}_a$, both at $p_\mathrm{ctrl}$. The data collapses in both panels with time rescaled as $t/L^z$, giving dynamical exponent $z=2.07(8)$ from $S_{1/2}$ and $z=1.98(2)$ from $\overline{S}_a$. We average over 1000 circuit realization in (a), 4000 in (b), and 1000 in (c).}
    \label{fig:ctrlplots}
\end{figure*}
The values of $\overline{S}_{1/2}$ and $\overline{S}_a$ at long times vary as functions of $p$, and a change in the qualitative behavior of the entropy can be used to identify a transition between a general, entangled (stabilizer) state and a state with no long-range entanglement. 
In the volume-law phase the half-cut entropy, by definition, scales with the volume of the system, here simply the system size $L$, while in the area law phase it saturates to an $O(1)$ value.
At the MIPT in Refs.~\cite{li_quantum_2018,skinner_measurement-induced_2019,vasseur_entanglement_2019}, the critical behavior of the half-cut entropy at long times manifests logarithmic scaling with system size, $\overline{S}_{1/2} \sim \alpha_L \ln(L)$---consistent with what we show in Sec.~\ref{sec:ent_transition} and what we see in \cref{fig:phases} but not what was found in the generic case \cite{Iadecola2023} when the control and entanglement transitions coincide.
The ancilla entropy $\overline{S}_a$ stays near its maximum value of $1$ (in units of $\ln 2$) in the volume-law phase and drops to $0$ in the area-law phase. 
In the thermodynamic limit the change from one behavior to the other occurs abruptly at the critical $p=p_\mathrm{ent}$, but for finite-sized systems the feature is broadened.
The transition can nevertheless be very precisely identified as the crossing point of $\overline{S}_a$ vs.~$p$ for different system sizes \cite{Gullans2020a}.
Finally, we note that the control transition in Ref.~\cite{Iadecola2023} itself was associated with a transition not to an area-law state, but to a \emph{disentangled} state; we find the same feature here except that for $p_\mathrm{ent} < p < p_\mathrm{ctrl}$ the system is now area-law entangled, see \cref{fig:phases}.

Additionally, we compute the expectation value of the magnetization density operator,
\begin{equation}
    \hat m = \frac1{L}\sum_{i=1}^L Z_i,
\end{equation}
where $Z_i$ is the Pauli-$Z$ operator on qubit $i$ and $Z_i \ket{0} = 1$.
This allows us to see whether the system has been controlled onto the ferromagnetic target state; the magnetization density will be $0$ on average in the uncontrolled phase since the qubits are randomly distributed and will reach its maximum value of $1$ when the target state is successfully prepared. 
This is the order parameter for the control transition.

\section{Control Transition}\label{sec:ctrl_transition}

\begin{figure*}[!ht]
    \includegraphics[width=0.42\textwidth]{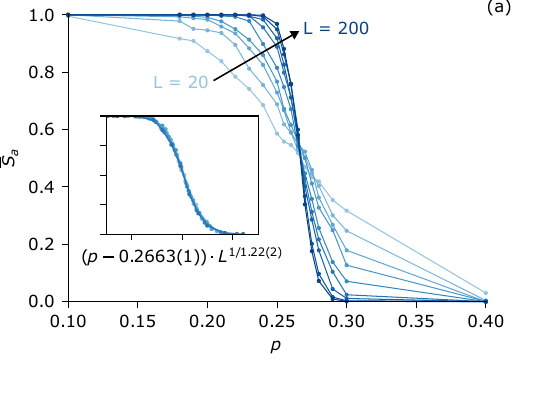}
    \includegraphics[width=0.42\textwidth]{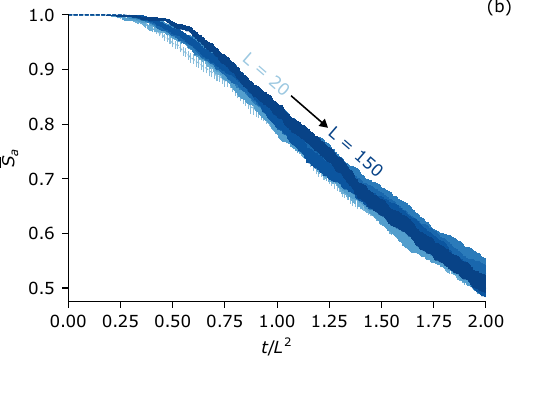} \\
    \includegraphics[width=0.42\textwidth]{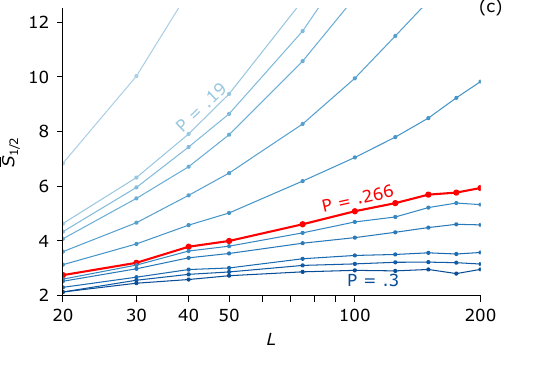}
    \includegraphics[width=0.42\textwidth]{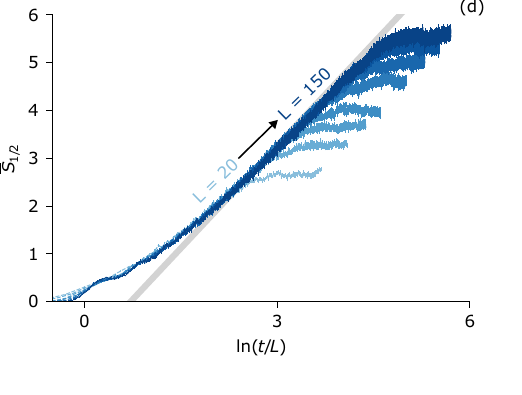}
     \caption{{\bf The Entanglement Transition}: (a)  The ancilla entropy $\overline{S}_a$ as a function of control probability $p$. A crossing is visible at $p_\mathrm{ent} = 0.266$, identifying the entanglement transition. Two-parameter data collapse, shown in the inset, gives $p_\mathrm{ent} = 0.2663(1)$ and $\nu = 1.22(2)$.
     (b) Dynamics of $\overline{S}_a$ at the entanglement transition $p_\mathrm{ent} = 0.266$. Collapse with time rescaled as $t/L^2$ should be understood as $T/L$ with $T=t/L$, giving $z\approx 1$. 
     (c) The half-cut entropy $\overline{S}_{1/2}$ as a function of system size $L$ for various control probabilities $p$. The thicker red line for $p = 0.266$ marks the transition between area-law and volume-law-entangled phases, where $\overline{S}_{1/2} \sim \alpha \ln(L)$ with $\alpha = 1.43(2)$.
     (d)  Dynamics of $\overline{S}_{1/2}$ at the entanglement transition $p_{\mathrm{ent}} = 0.266$. The collapse as a function of $\ln(T)$ demonstrates that $T=t/L$ is the natural time unit for entanglement dynamics at this transition. Agreement with the grey guide line of the form $\overline{S}_{1/2} \sim \alpha \ln(t/L)$, using $\alpha = 1.43$ as extracted from (c), demonstrates that $z\approx 1$ at this critical point. All panels show averages over 1000 circuit realizations.}
     \label{fig:MIPTplots}
\end{figure*}

We first analyze the transition between uncontrolled and controlled phases. 
In prior work~\cite{Iadecola2023} the transition between volume-law entangled and disentangled phases occurred at the same point as the transition between uncontrolled and controlled phases (to within numerical accuracy), but here we find that the use of stabilizers strongly splits the two apart.
As described in Sec.~\ref{sec:model}, we can pinpoint the control transition by computing the magnetization density at late times.
In \cref{fig:ctrlplots}(a) the magnetization density at $t=2L^2$ is plotted against $p$ for a range of system sizes. We find a clear signature of the transition around the known location of the classical control transition, $p_\mathrm{ctrl}= 0.5001(1)$. 
With the transition point identified, data collapse [see inset of \cref{fig:ctrlplots}(a)] allows us to extract the correlation length critical exponent $\nu = 1.00(1)$, also in agreement with previous results~\cite{antoniou_absolute_1998,Iadecola2023}.

Though we are well outside the volume-law-entangled phase at $p_\mathrm{ctrl}=0.5 > p_\mathrm{ent}$ we can still extract useful information about the nature of the transition by analyzing how entanglement entropy changes with time.
In \cref{fig:ctrlplots}(b) and (c) we show the dynamics of $\overline{S}_{1/2}$ and $\overline{S}_a$ and find that these quantities collapse upon rescaling $t\to t/L^z$ with $z\approx2$ in both cases---we find $z = 2.07(8)$ and $1.98(2)$ from the half-cut and ancilla entropies, respectively.
This is consistent with the expectation from the classical transition~\cite{antoniou_absolute_1998,Iadecola2023}; the domain wall between controlled and uncontrolled regions of the system can be viewed classically as executing a random walk on the scale of the time steps measured by $t$, with control and Bernoulli steps pushing it in opposing directions.
Further, the half-cut entanglement growth with $\overline{S}_{1/2}\sim\sqrt{t}$ in \cref{fig:ctrlplots}(b) is consistent with the results of previous work~\cite{Iadecola2023} for the generic (Haar) case, which were obtained at much smaller system sizes.
Furthermore, we see that at the transition the half-cut entanglement saturates to $\overline{S}_{1/2}\sim 0.172(7)$ (error computed via standard deviation of $\overline{S}_{1/2}(t)$ for late times and largest system size), different from the generic case where it saturated to $\overline{S}_H \sim 0.75(5)$, an indication of different quantum fluctuations at the control transition.

Last, while the order parameter $\hat{m}$ can observe this control transition, the measurement record from the $R_L$ operation can as well (see Appendix~\ref{app:record}); however, universal scaling cannot as easily be extracted from the record.

\section{Entanglement Transition}\label{sec:ent_transition}

We now turn to the entanglement entropy to identify a transition between volume-law and area-law entangled phases.
As discussed above, we calculate $\overline{S}_{1/2}$ and $\overline{S}_a$ from the stabilizer tableau after stochastically applying the Bernoulli and control maps for $2L^2$ time steps. 
In \cref{fig:MIPTplots}(a) we show the ancilla entanglement entropy for various system sizes at $t=2L^2$, and the crossing point indicates the presence of a transition. 
With a two-parameter data collapse of $\overline{S}_a$, shown in the inset of \cref{fig:MIPTplots}(a), we are able to extract $p_\mathrm{ent} = 0.2663(1)$ and the correlation length critical exponent $\nu = 1.22(2)$. 
The critical value $\nu$ differs from the concomitant entanglement-control transition found in Ref.~\cite{Iadecola2023} and instead agrees well with the values found for the Clifford MIPT~\cite{gullans_scalable_2020,zabalo_critical_2020}.

To confirm other features of the Clifford MIPT, we compute the half-cut entanglement entropy at $t=2L^2$ and for $p$ values around $p_\mathrm{ent}$ as a function of $L$, shown in \cref{fig:MIPTplots}(c).
We find the expected linear growth with $L$ for $p<p_\mathrm{ent}$ (volume-law) and saturation to an $O(1)$ constant for $p>p_\mathrm{ent}$ (area-law). 
At criticality ($p = p_\mathrm{ent}$), the data fits the expected logarithmic form, $\overline{S}_{1/2} \sim \alpha \ln(L)$ with $\alpha = 1.43(2)$.
Notably, this differs from the Clifford MIPT value of $\alpha_C = 1.61(3)$ \cite{zabalo_critical_2020}.
Interestingly, this implies that the control operation has made the system less entangled at the critical point compared to the conventional Clifford MIPT.

We also determine the dynamical critical exponent $z$ at $p_\mathrm{ent}$, which will help further specify the universality class at the critical point. 
To do so we analyze the time-dependence of the entanglement entropy, however the circuit time $t$---the value that increments by 1 whenever we apply a Bernoulli or control map---is not the most natural choice of time for this analysis. 
To understand why, we appeal to universality at the critical point:
the dynamical exponent relates space and time via $T \sim L^z$, and we can therefore relate the value of $\overline{S}_{1/2}$ at late times to $\overline{S}_{1/2}$ at early times (when entanglement is still growing) by replacing $L$ with $T^{1/z}$.
In this language, at $p = p_\mathrm{ent}$
\begin{equation}
    \overline{S}_{1/2} \sim \begin{cases}
     \alpha \ln L, & T \gg L, \\
     \frac{\alpha}{z} \ln T, & T \ll L.
     \end{cases}
\end{equation}
However, we observe that the entanglement dynamics in our simulations are not $L$-independent unless we assume $T = t/L$ as we demonstrate in \cref{fig:MIPTplots}(d).

This observation can be understood in terms of the entanglement density generated by a single time step.
One application of the Bernoulli map applies a single entangling gate, so to increase the half-cut entropy by a single unit requires $O(L)$ time steps.
For comparison, a brickwork circuit does this in a \emph{single} time step.
Therefore instead of the bare circuit time $t$, the natural unit of time for entanglement dynamics near $p_\mathrm{ent}$ is the rescaled time $T=t/L$. The dynamical exponent $z$ should therefore relate space and \emph{this} notion of time through $T = L^z$.
Returning to \cref{fig:MIPTplots}(d), the time-dependence of $\overline{S}_{1/2}$ at $p_\mathrm{ent}$ is seen to collapse well with coefficient $\alpha/z = 1.348(2)$.
Recalling that we found $\alpha = 1.43(2)$ above, we therefore obtain $z = 1.06(2)$ from the half-cut entropy. 
For comparison, \cref{fig:MIPTplots}(b) shows the dynamics of the ancilla entropy $S_a$ at $p=p_\mathrm{ent}$.
The collapse of the data in this figure with $t/L^2$ should be understood in light of the above argument as a collapse in $T/L$, and doing so we extract $z=1.002(6)$, in good agreement with our other estimate.
These critical data suggest we have recovered the Clifford MIPT except for the coefficient $\alpha$ of the logarithmic growth of the half-cut entropy, which appears to differ by an amount $\alpha_C - \alpha = 0.18(4)$.

\begin{table}
    \centering
    \begin{ruledtabular}
    \begin{tabular}{l d{1.5} d{1.5} d{1.4} d{1.3}}
        & \mc{Entanglement} & \mc{Control} & \mc{Clifford \cite{zabalo_critical_2020}} & \mc{B-H \cite{Iadecola2023}}\\
        \hline
        $p_c$ & 0.266(3) & 0.5001(1) & 0.154(4) & 0.51(1) \\ 
        $\nu$ & 1.22(2) & 1.00(1) & 1.24(7) &  0.9(1) \\
        $z$, half-cut & 1.06(2) & 2.07(8) & \mc{$-$} &  \mc{$-$} \\
        $z$, ancilla & 1.002(6) & 1.98(2) & 1.06(4) &  2.1(1) \\
        $\alpha$ & 1.43(2) & \mc{$-$} & 1.61(3) & \mc{$-$} \\
        $\overline{S}_{1/2}$ & \mc{$-$} & 0.172(7) & \mc{$-$} & 0.75(5)
    \end{tabular}
    \caption{The numerically determined values of the the critical probabilities for the two transitions, the critical exponents $\nu$ and $z$, the coefficient of the logarithmic behavior of $\overline{S}_{1/2}$ at the critical point $\alpha$, and the value $\overline{S}_{1/2}$ saturates to at long times when finite.
    We also give the corresponding values for the Clifford MIPT~\cite{zabalo_critical_2020} and the Bernoulli circuit with Haar scrambling gates (B-H)~\cite{Iadecola2023}. 
    At the entanglement transition $\nu$ is determined from $\overline{S}_a$, and at the control transition it is determined from $\overline{\expval{\hat{m}}}$.
    The dynamical exponent $z$ is determined from both $\overline{S}_{1/2}$ and $\overline{S}_a$ at both critical points.}
    \label{tab:numbers}
    \end{ruledtabular}
\end{table}

\section{Discussion}\label{sec:discussion}

Remarkably, the introduction of Clifford scrambler gates (as opposed to generic Haar scramblers as used in Ref.~\cite{Iadecola2023}) has naturally separated the entanglement and control transitions.
This implies an intermediate area-law and uncontrolled phase as shown in \cref{fig:phases}.
We can begin to understand this with the first domain wall picture introduced in Eq.~\eqref{eq:fdw}.
The first domain wall explains why the control transition is governed by a random walk universality ($\ell$ randomly increases or decreases with probability $p$ and $1-p$, respectively).
However, when $p_\mathrm{ent} \leq p < p_\mathrm{ctrl}$, then $\ell \approx 0$ in the steady state.
In this case, the physics is governed by $\ket{\psi}$ in Eq.~\eqref{eq:fdw} which in the generic Haar case is volume-law entangled up until $p_\mathrm{ctrl}$.
With Clifford gates, on the other hand, $\ket{\psi}$ is a stabilizer state undergoing hybrid Clifford dynamics.
Therefore, it is plausible that $\ket{\psi}$ could undergo its own volume-law to area-law transition as we have observed.
This appears to be due to the finite probability for a Clifford gate to not introduce any entanglement or even to remove entanglement~\footnote{As an example, there are 4 single-qubit stabilizer states and 24 two-qubit stabilizer states~\cite{Gross2006a}. Since we can enumerate both disentangled and entangled stabilizer states, there is a nonzero chance that any given Clifford gate can disentangle an entangled stabilizer state.}---a feature the generic Haar gate does not have~\cite{Page1993}.

The hybrid Clifford dynamics $\ket{\psi}$ experiences is in detail different from \cite{li_quantum_2018,li_measurement-driven_2019,Gullans2020a}, but nonetheless appears to flow to a similar universality class (see \cref{tab:numbers}).
The exception, as noted in Section~\ref{sec:ent_transition}, is the coefficient of the logarithm $\overline{S}_{1/2}\sim \alpha \ln L$.
Furthermore, this universality only revealed itself once it was clear the system had a natural time-step at $p_\mathrm{ent}$ (enumerated by $T = t/L$) distinct from the time-step dictating the random walk of the first domain wall (enumerated by $t$).
We leave it to future work to determine additional properties of the universality class such as the critical exponent $\eta$ \cite{zabalo_critical_2020} and the effective central charge $c_\mathrm{eff}$ \cite{zabalo_operator_2022}.

Last, the control transition appears to be largely unaltered from either the classical version or the generic quantum version.
This suggests that the control transition becomes the dominant physics while the entanglement transition could have a different universality.
This is suggested by the results in Ref.~\cite{Iadecola2023} where the entanglement and control transitions coincide; in that case, all critical properties appear to be governed by the control transition.
In the stabilizer model, the Clifford gates do not provide enough entanglement to move these transitions together (not shown), but other formulations of these dynamics could make this manifest.
Nonetheless, we have shown that in the stabilizer Bernoulli map $p_\mathrm{ent}$ occurs separate from $p_\mathrm{ctrl}$ and with its own, distinct, universality.

\begin{acknowledgements}
We thank Sriram Ganeshan for valuable discussions and collaboration on related work and Ian Spielman for discussions leading to the ideas in the appendix.  
This work was supported in part by the National Science Foundation under Grants No.~DMR-2238895 (C.~L. \& J.~H.~W.) and No.~DMR-2143635 (T.~I.), the Office of Naval Research grant No.~N00014-23-1-2357 (J.H.P.), and the Alfred P. Sloan Foundation through a Sloan Research Fellowship (J.H.P.). 
This work was initiated and performed in part at the Aspen Center for Physics, which is supported by the National Science Foundation Grant No.~PHY-1607611. 
Portions of this research were conducted with high performance computational resources provided by Louisiana State University (http://www.hpc.lsu.edu).
\end{acknowledgements}

\appendix
\section{Measurement Record}\label{app:record}

\begin{figure}[!ht]
    \centering
    \includegraphics[width=\columnwidth]{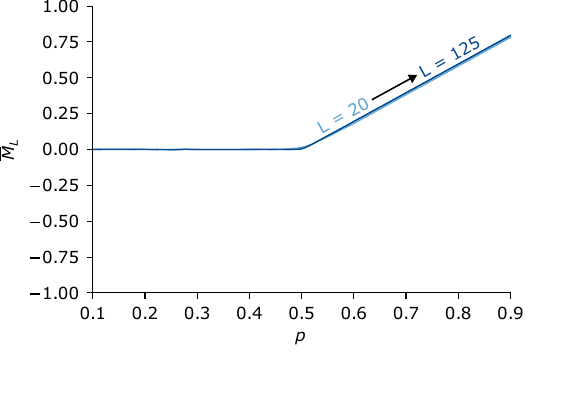}
    \caption{The average of the measurement record as a function of $p$ for $L = 20, 30, 40, 50, 75, 100, 125$. In addition to averaging all measurement outcomes in a given realization of the circuit, this result also averages over the outcomes of 1000 realizations.}
    \label{fig:Bernoulli_Measure}
\end{figure}

During the application of the Bernoulli and control maps the reset as in \cref{eq:Cst} measures qubit $L$. 
Analyzing the record of these measurement outcomes gives insight into the control transition and helps quantify the measurement trend in the controlled and uncontrolled phases. 
We show the average values of this record in \cref{fig:Bernoulli_Measure} as a function of $p$, where entries in the record are formatted as a $+1$ for measurement outcome $0$ and a $-1$ for measurement outcome $1$.
In the limit of infinite system size, for $p<p_\mathrm{ctrl}$ the average over the measurement record is therefore $0$, while for $p>p_\mathrm{ctrl}$ it increases linearly from $0$ to $1$,
\begin{equation} \label{eq:MLavg}
    \overline{M}_L = \begin{cases}
        0, & p<p_\mathrm{ctrl}, \\
        2p-1, & p>p_\mathrm{ctrl}.
    \end{cases}
\end{equation}
The behavior in the uncontrolled regime is easily understood.
The final qubit is always randomly scrambled since there are more Bernoulli steps than control steps, so measurements on qubit $L$ during the reset operation of the control steps are always equally likely to be $\pm1$ and the average is $0$.

Understanding the behavior in the controlled regime, despite the simple form, is less straightforward.
First note that when the system is in the target state $\ket{00\cdots0}$, application of a single Bernoulli map scrambles the final $s$ qubits if we consider an $s$-qubit scrambling gate, but further applications scramble only one additional qubit at a time.
(In the main text we use an $s=2$ scrambling gate $S_\mathrm{st}$ in the implementation of the Bernoulli map \cref{eq:Bst}, also shown in \cref{fig:schematic}, but will leave this number general for the moment.)
Therefore, after applying the Bernoulli map $N_B$ times to the target state, there are $N_B+s-1$ scrambled qubits, and as many subsequent sequential control steps are needed to return the system to the target state, which would yield measurements that average to $0$.
This remains true if the Bernoulli and control maps are reordered as long as the system is not put into the target state until the final step. 
Additional control steps after this point are guaranteed to measure $1$. 
In any circuit, the record of maps can generically be divided into segments following this pattern---$B_\mathrm{st}$ and $C_\mathrm{st}$ are applied $N_B$ and $N_B+s-1$ times (in some order) to take the system from the fully controlled state to itself, yielding $N_B+s-1$ measurements with $0$ average, then $C_\mathrm{st}$ is applied an additional $N_1$ times, measuring $1$ each time.
In total there are $N_C = N_B + N_1 + s - 1$ applications of $C_\mathrm{st}$, and the average measurement is $N_1/N_C$ for one such segment of the entire circuit.

In each of these segments of $N = N_B + N_C$ time steps $N_B$ and $N_C$ will vary, but for long times this variation will average out and we can calculate the average measurement by determining \emph{average} numbers of applied gates.
We can start by putting
\begin{gather}
    \overline{N}_B = (1-p)\overline{N} \label{eq:NBavg}\\
    \overline{N}_C = \overline{N}_B + \overline{N}_1 + s - 1 = p\overline{N}, \label{eq:NCavg}
\end{gather}
so the average number of Bernoulli and control gates are exactly related to the average number of total gates by the expected probabilities.
We can determine $\overline{N}_1$ explicitly---it is the expected number of times we sequentially apply $C_\mathrm{st}$ once we arrive back at the target state.
Since the control probability is $p$, we have
\begin{equation} \label{eq:N1avg}
    \overline{N}_1 = \frac{\sum_{\ell=0}^\infty \ell p^\ell}{\sum_{\ell=0}^\infty p^\ell} = \frac{p}{1-p}.
\end{equation}
Combining \cref{eq:NBavg,eq:NCavg,eq:N1avg} we can find the average total number of gates,
\begin{multline}
    (1-p)\overline{N} + \frac{p}{(1-p)} + s-1 = p\overline{N} \\
    \Rightarrow \quad \overline{N} = \frac{p + (s-1)(1-p)}{(2p-1)(1-p)}.
\end{multline}
The measurement average is just the fraction of measurements that are guaranteed to find $1$,
\begin{equation}
    \overline{M}_L = \frac{\overline{N}_1}{\overline{N}_C} = \frac{(2p-1)}{p+(s-1)(1-p)},
\end{equation}
which for $s=2$ gives the result in \cref{eq:MLavg}.

Note that for scrambling gates of different sizes $s$ the behavior of the average over the measurement record will change drastically; for $s$ large, $\overline{M}_L$ stays near $0$ for $p>p_\mathrm{ctrl}$ even though the system is typically controlled very near to the target state.
This is simply because the local information obtained from measuring only of the final qubit is not necessarily a good proxy for the global properties of the system. 

\bibliography{refs, zotero_refs}

\end{document}